\documentclass[aps,prl,groupedaddress,fleqn,twocolumn]{revtex4-1}
\pdfoutput=1
\usepackage{graphicx}
\usepackage{amsmath}
\usepackage{amssymb}
\usepackage{gensymb}

\newcommand\alphavalue{0.28\pm 0.04}

\begin{document}

\title{Linewidth enhancement factor measurement based on FM-modulated optical injection: application to rare-earth doped active medium}

\author{Aur\'elien Thorette$^*$}
\author{Marco Romanelli}
\author{Marc Vallet}

\address{\small Institut de Physique de Rennes, Universit\'e Rennes I - CNRS UMR 6251 \\
	Campus de Beaulieu, 35042 Rennes Cedex, France	}
\address{$^*$Corresponding author: aurelien.thorette@univ-rennes1.fr}

\begin{abstract}
A new method for measuring the linewidth enhancement factor of a laser is proposed. It is based on frequency-modulated optical injection, combined with dual-frequency laser operation. The linewidth enhancement factor $\alpha$ is deduced from the experimental data using a theoretical analysis based on a standard rate equation model. As the intracavity power is kept constant, the method allows to free the process from the thermal effects that are usually present in AM/FM techniques. Measurement of $\alpha=\alphavalue$ in a diode-pumped Nd:YAG laser demonstrates that the method is well-suited for characterizing small values of $\alpha$.
\end{abstract}

\maketitle
	
The linewidth enhancement factor, also referred to as {\em Henry} factor or $\alpha$ factor~\cite{henry1982}, quantifies the phase-amplitude coupling in a laser gain medium. The origin of $\alpha$ comes from an asymmetric gain profile or from a detuning of the laser frequency with respect to the gain line center. In semiconductors, for which $\alpha$ can take rather large values, this phase-amplitude coupling describes important characteristics of laser behavior, such as a large broadening of the laser linewidth~\cite{petermann2012laser} or peculiar dynamics under current modulation or optical injection~\cite{wieczorek2005}. While it is fairly common to consider $\alpha\simeq0$ for active media with more symmetric gain profiles such as diode-pumped solid-state lasers, a small phase-amplitude coupling may also have to be taken into account when targeting applications needing stabilized solid state lasers with very low optical phase noise, such as gravitation wave detection~\cite{acernese2009,abbott2016} or optically carried radiofrequency generation~\cite{pillet2008}.

\par Very extensive literature exist on $\alpha$-factor measurements performed in all main types of semiconductor lasers, i.e., quantum cascade lasers, quantum dots, VCSELs and so forth. The measurement methods include direct estimation of the gain asymmetry~\cite{hakki1975}, pump induced phase modulation through AM/FM coupling~\cite{provost2011} and optical injection~\cite{hui1990,iiyama1992}. Conversely, studies of the phase-amplitude coupling in solid-state lasers have been much less common. A value of $\alpha = 0.25\pm 0.13$ has been found in a Nd:YVO4 laser, using either injection~\cite{valling2005} or pump AM/FM modulation method~\cite{fordell2005}, while a surprisingly large $\alpha\approx1$ was reported in Nd:YAG microchip lasers~\cite{szwaj2004}.

\par Any measurement of $\alpha$ needs a way to either force, as in injection methods~\cite{valling2005}, or measure, as in AM/FM modulation~\cite{fordell2005}, the optical phase. In both cases, one needs to precisely control the optical frequency difference between the laser under study and an auxiliary optical source. Here, we propose to use a laser operating in a dual-frequency regime, thus providing simultaneously the master and the slave oscillator. In this way, we can take advantage of the intrinsic stability of the frequency difference between the modes, and of their perfect mode-matching (both due to the fact that they share the same optical cavity) ~\cite{baxter1996,brunel2004}. However, while dual-frequency operation facilitates the implementation of the method, we stress that the latter does not require it, and could be equally used in the standard optical injection configuration. 

\par The aim of this letter is thus to present an ``FM/AM'' injection method based on the amplitude response of a lasing mode to a frequency-modulated optical injection of a second mode, and to show how its implementation in a Nd:YAG dual-frequency laser leads to a rather precise characterization of small $\alpha$ factors. 
\\
\par For the sake of clarity, we first describe the method on an ideal master-slave injection configuration. The rate equation for the electric field of an  injected class-B laser is generically~\cite{wieczorek2005} :

\begin{equation}
	\label{eq:injection}
	\frac{d\mathcal{E}}{dt} = (1+i\alpha)\frac{N\mathcal{E}}{2}+i\Delta\mathcal{E} + \Gamma E_\mathrm{inj}
\end{equation}

Here, $\mathcal{E}$ is the intracavity field, $N$ the active medium gain, $\Delta$ is the detuning between the injected field and the free-running laser frequency, $\Gamma$ is the injection efficiency, and $E_\mathrm{inj}$ the injected field, whose phase is taken as reference. Separating phase and amplitude as $\mathcal{E} = E\exp(i\varphi)$ leads to : 

\begin{subequations}
	\label{eq:injectionphase}
	\begin{align}
		\label{eq:injection1}
		\frac{dE}{dt}  &= \frac{NE}{2} + \Gamma E_\mathrm{inj} \cos\varphi \\
		\label{eq:injection2}
		\frac{d\varphi}{dt}  &= \alpha\frac{N}{2} + \Delta - \Gamma \frac{E_\mathrm{inj}}{E}\sin\varphi
	\end{align}
\end{subequations}


We consider small perturbations of the injection-locked, steady state regime. Thus, we write $x = \widehat{x}+\delta x$, where $x$ stands for ${E,\varphi,N}$. $\widehat{x}$ denotes the steady state value of $x$ and $\delta x$ the small perturbation. Linearization of equation (\ref{eq:injection1}) leads to :
\begin{equation}
	\frac{d\delta E}{dt} = \frac{\widehat{E} \delta N + \widehat{N} \delta E}{2} - \Gamma E_\mathrm{inj} \sin\widehat{\varphi}\delta\varphi
\end{equation}
This shows clearly that amplitude response to a phase perturbation $\delta\varphi$ depends on the quantity $\sin\widehat{\varphi}$. In particular, a zero response is expected when $\sin\widehat{\varphi}=0$. Using the steady state equation (\ref{eq:injection2}), this condition becomes $\alpha \widehat{N}/2 = - \Delta$, which we can transform using  (\ref{eq:injection1}) to the more useful expression :
\begin{equation}
	\label{eq:delta_m_inj}
	\Delta = \alpha\Gamma\frac{E_\mathrm{inj}}{\widehat{E}}
\end{equation}
\par This detuning corresponds to a minimal amplitude response to a phase perturbation, and this result shows that it is directly related to $\alpha$. Consequently, it provides a way to measure phase to amplitude coupling, and will be at the root of our method. In the following, we will denote this value as the minimal amplitude response detuning $\Delta_m$. We point here that this method is only suited to small values $\alpha < 1$, because it relies on the measurement of $\Gamma$, which can only be derived from the span of the injection locking region $\Delta^+-\Delta^-$. This region is roughly $|\Delta|<\Gamma\sqrt{1+\alpha^2} E_\mathrm{inj}/\widehat{E}$ for low injection. Thus we have $\alpha/\sqrt{1+\alpha^2}=2\Delta_m/(\Delta^+-\Delta^-)$, where the left-hand term only has a dependance in $\alpha^{-2}$ for high values $\alpha > 1$, making any precise measurement impractical.  On the other hand, for $\alpha < 1$, it scales as $\alpha$, which makes this method well adapted to the low values expected for solid-state lasers.

\par
\begin{figure}[htbp]
	\centering
	\includegraphics{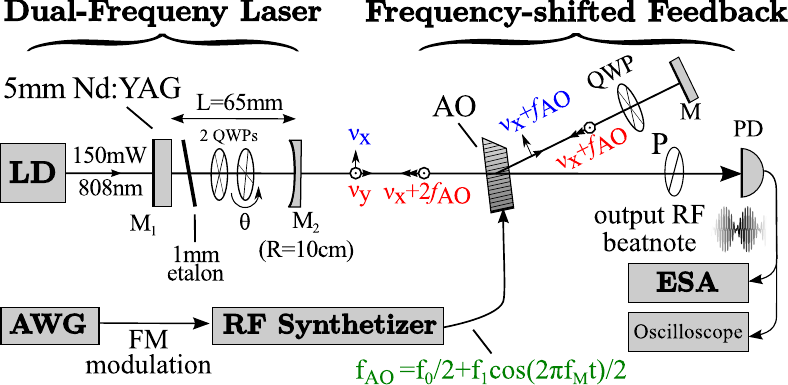}
	\caption{Experimental setup. An intracavity birefringent element (here, two rotated quarter-wave plates QWPs) forces the laser to produce two polarization modes. One of them is frequency-shifted with an acousto-optic modulator (AO) close to the other one, and injected back in the cavity. The output signal is an optically-carried RF beat-note. See text for details.}
	\label{fig:setup}
\end{figure}

 As for all injection methods, the measurement requires a very stable and tunable master laser. In order to bypass this requirement, we now discuss the alternative solution in which we will use the laser as its own reference, by making it dual-frequency, and using one of the modes to inject on the other through a feedback. The experimental setup is summarized in Fig.~\ref{fig:setup}, which is similar to the one described in ~\cite{romanelli2014}. It is centered around a $\mathrm{5mm}$-long (111)-cut Nd\textsuperscript{3+}:YAG crystal operating at $\mathrm{1064nm}$. The crystal is optically pumped with a 150mW single mode circularly polarized laser diode LD at 808nm. The associated cavity is $L=\mathrm{6.5cm}$ long. One  high-reflection mirror ($M_{1}$) is directly coated on the external face of the crystal and the cavity ends with a 10cm radius mirror ($M_{2}$) with 99\% reflectivity. In order to achieve dual-frequency operation, a tunable phase retardance is added in the cavity by using two quarter-wave plates (QWP), of which one is rotated by an angle $\theta$. This induces an optical path difference for the two polarization eigenmodes, which leads to a lasing frequency difference $\delta\nu=\nu_y-\nu_x=\theta c/\pi L$. A 1mm-thick etalon is also inserted to ensure single longitudinal mode operation for each polarization state. Dual polarization, dual-frequency operation can thus be achieved, the frequency offset between the two modes being tunable from 0 up to $c/4L$. Here, we choose $\delta\nu\approx 200\mathrm{MHz}$. The two output modes are combined by a polarizer P. The resulting RF beatnote is the useful signal, which is monitored by a photodiode PD followed by an electric spectrum analyser ESA and on an oscilloscope.

\par We then simply inject one mode in the other using a feedback external cavity. An acousto-optic modulator (AO) driven at a frequency $f_\mathrm{AO}$ is used to shift the optical frequencies so that $\nu_x+2f_\mathrm{AO}\approx\nu_y$. Then a mirror M and a QWP are used to inject back the $x$-polarization in the $y$-polarization in the laser cavity. The intensity of this reinjection can be controlled by the diffraction efficiency of the modulator, and the detuning between the injected field and the intracavity field is then $\delta = \delta\nu-2f_\mathrm{AO}$. When this detuning is kept small, stable phase locking occurs between the output beatnote and the RF synthetizer. The size of this locking range depends on the feedback efficiency, and can extend to 2MHz.

\par We then introduce a phase perturbation through a modulation of the detuning $\delta$. This can be conveniently achieved using FM modulation of the RF synthetizer that drives the acousto-optic modulator. This leads to a frequency shift of $2f_\mathrm{AO} = f_0+ f_1\cos(2\pi f_M t))$. We choose to modulate the detuning at a frequency $f_M$ close of the relaxation oscillations frequency $f_R$ of the laser that we measure to be $f_R=\mathrm{60kHz}$, so that the AM response is maximized by the resonance. As the perturbation has to be kept small, we take $f_1=10\mathrm{kHz}$. 

\par
\begin{figure}[htbp]
	\centering
	\includegraphics{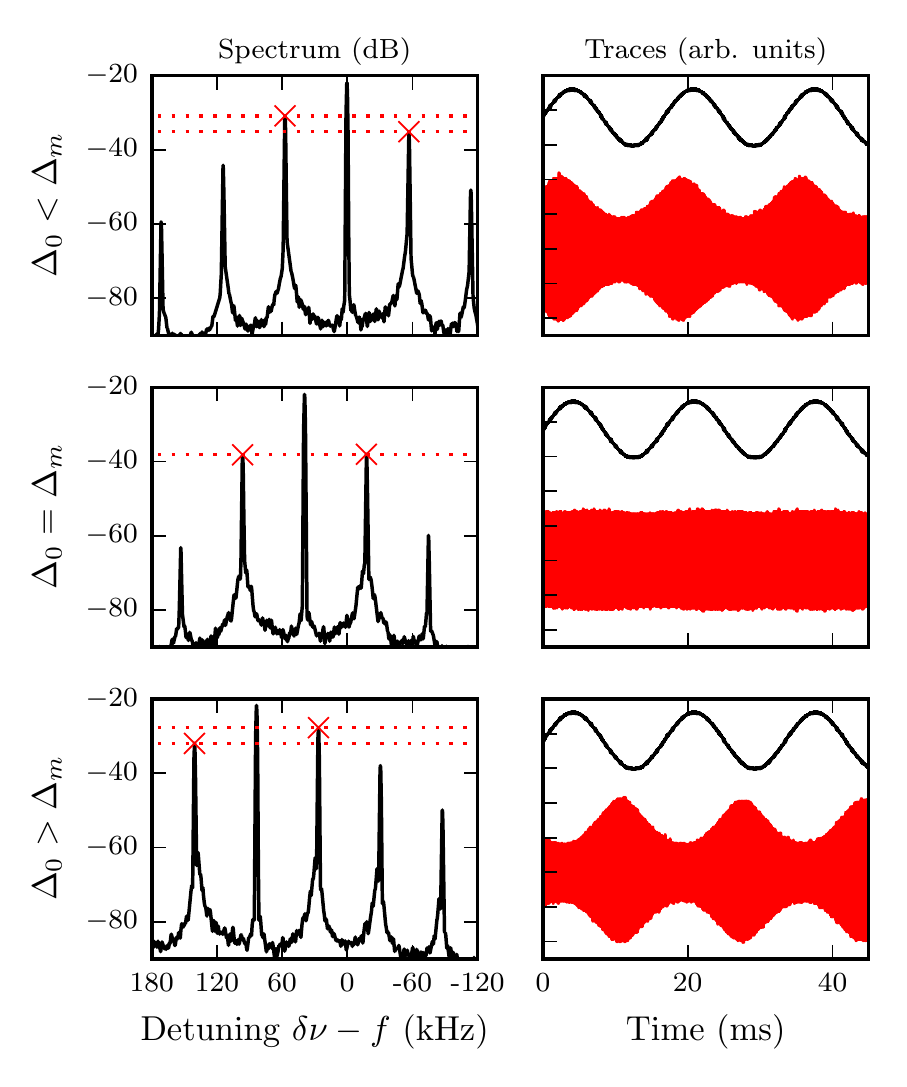}
	\caption{Power spectrum for $\Delta_0<\Delta_m$, $\Delta_0=\Delta_m$ and $\Delta_0>\Delta_m$ (with $\Delta_0=(\delta\nu-f_0)/f_R$), and the associated time series (black: modulation signal, red: output beatnote $I_{xy} = |e_x+e_y|^2$). This shows that the balance of the two sidebands at $\pm f_R$ corresponds to minimal amplitude response, and to $\pi$ phase shift between $\Delta_0>\Delta_m$ and $\Delta_0<\Delta_m$}
	\label{fig:spectrum}
\end{figure}

\begin{figure}[htbp]
	\centering
	\includegraphics{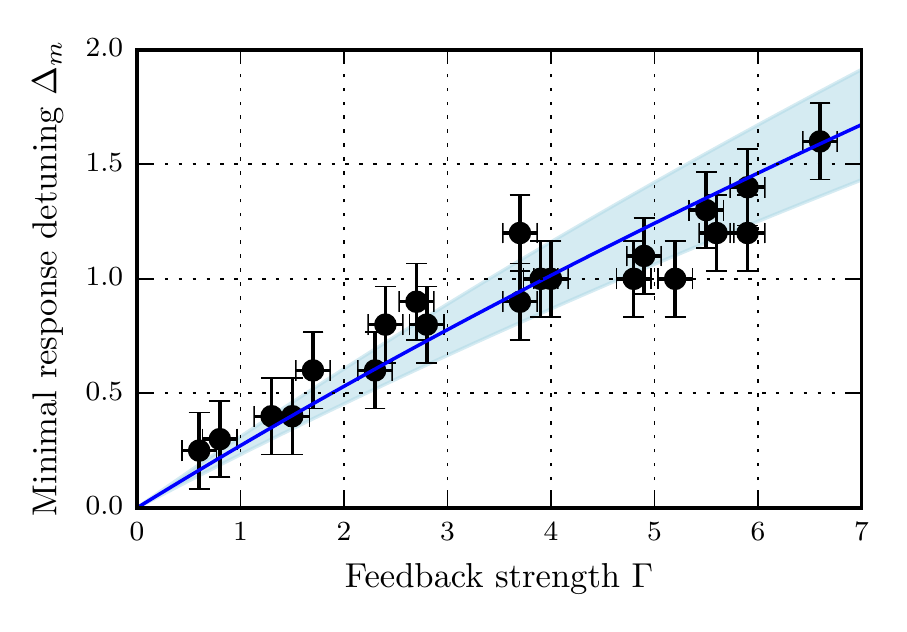}
	\caption{Experimental values of $\Delta_m$ (normalized to the relaxation oscillation frequency $f_R$ = 60 kHz). Blue curve: best fit from equation~(\ref{eq:delta_m_comp}), corresponding to $\alpha=\alphavalue$. The shaded region shows the computed values of $\Delta_m$ within the error range for $\alpha$.}
	\label{fig:result}
\end{figure}

 We have introduced a phase modulation and are now interested in the resulting amplitude modulation, which can be related to the expected phase-amplitude coupling $\alpha$. We observe that the amplitude modulation response depends strongly on the  mean detuning $\delta\nu-f_0$, as shown in Fig.~\ref{fig:spectrum}. With no phase-amplitude coupling, the amplitude response would be minimal for a null detuning, as implied by equation (\ref{eq:delta_m_inj}). But experimentally we observe that the amplitude modulation is minimal for a positive value of the detuning, suggesting a measurable non-zero value of $\alpha$. We notice that the minimal amplitude response corresponds to an equal intensity for the two side peaks at $f_0\pm f_M$. This balance criterion between the two sidebands is most convenient for an efficient measurement, as shown by Fig.~\ref{fig:spectrum}. Furthermore, Eq.~\ref{eq:delta_m_inj} implies that $\Delta_m$ should increase with the injection strength. We have thus repeated the measurement for increasing values of the reinjection efficiency, by varying the efficiency of the acousto-optic diffraction. The results are plotted in Fig.~\ref{fig:result}, and show indeed that $\Delta_m$ depends strongly on the injection strength, again indicating a non-zero value of the phase-amplitude coupling.

\par In order to have a quantitative comparison between the experiments and the theory, and to extract a value of $\alpha$, we have to refine the model of equations (\ref{eq:injection}) to include some features that are specific of dual-frequency operation (i.e. the coupling between the two eigenmodes in the active medium). We can model our laser using the following rate equations for the normalized fields ($e_{x,y}$) and the normalized population inversions ($m_{x,y}$)~\cite{thevenin2012a}, to which we add the phase-amplitude coupling as an imaginary part for the gain:


\begin{subequations}
	\label{eq:model}
	\begin{align}
		\label{eq:model1}
		\frac{de_x}{ds}  =& \,(1+i\alpha)\frac{m_x+\beta m_y}{1+\beta}\frac{e_x}{2} \\
		\label{eq:model2}
		\frac{de_y}{ds}  =& \,(1+i\alpha)\frac{m_y+\beta m_x}{1+\beta}\frac{e_y}{2} + i\Delta e_y + \Gamma e_x \\
		\begin{split}
			\label{eq:model3}
			\frac{dm_{x,y}}{ds} =& \, 1- (|e_{x,y}|^2+\beta |e_{y,x}|^2) \\ &- \varepsilon m_{x,y} [1+(\eta-1)(|e_{x,y}|^2+\beta |e_{y,x}|^2)] 
		\end{split}
	\end{align}
\end{subequations}

The phase of the field complex electric field $\mathcal{E}_x$ is taken as a reference, and we use $\mathcal{E}_x = e_x \exp(2i\pi \nu_x t)$ and $\mathcal{E}_y = e_y \exp(2i\pi(\nu_x+2 f_{AO})t)$. The time scale $s=2\pi f_R t$ is related to the relaxation oscillations. Control parameters are the normalized feedback intensity $\Gamma$ and the normalized detuning $\Delta = (\delta\nu-2f_\mathrm{AO})/f_R$. Constant parameters are pump rate $\eta$, the coupling inside gain medium $\beta$, and $\varepsilon= \gamma_{\parallel}/2\pi f_R$ that accounts for the populations lifetime $1/\gamma_\parallel$. All these parameters can be measured or experimentally controlled. The mode coupling $\beta$ can be infered from the frequency of antiphase oscillations~\cite{lacot1996}, which are observed by monitoring the intensity noise of a single polarization mode. We measure them to be at a frequency $f_A$ so that $\Omega=f_A/f_R=0.66$, so we have $\beta=(1-\Omega)/(1+ \Omega)=0.20\pm0.05$.
Well known values of $1/\gamma_{\parallel}=230\mu s$ leads to $\varepsilon\approx0.01$~\cite{brunel1999}.
$\eta$ is the ratio of the pump laser diode current to the threshold current, and we have used a value of $1.2\pm0.1$.
The least known parameter here is $\Gamma$, which is related to the intensity injected from one mode to the other. 
Since it has been shown that phase locking between the laser beatnote and the RF synthetizer occurs when $|\Delta|<\Gamma$  and that it is characterized by a single peak at $2f_\mathrm{AO}$ on the RF power spectrum, $\Gamma$ can be obtained by measuring the phase-locking range. The experimental beat-note signal can be compared with the computed quantity $I_{xy} = |e_x+e_y|^2$.
\\


\begin{figure}[htbp]
	\centering
	\includegraphics{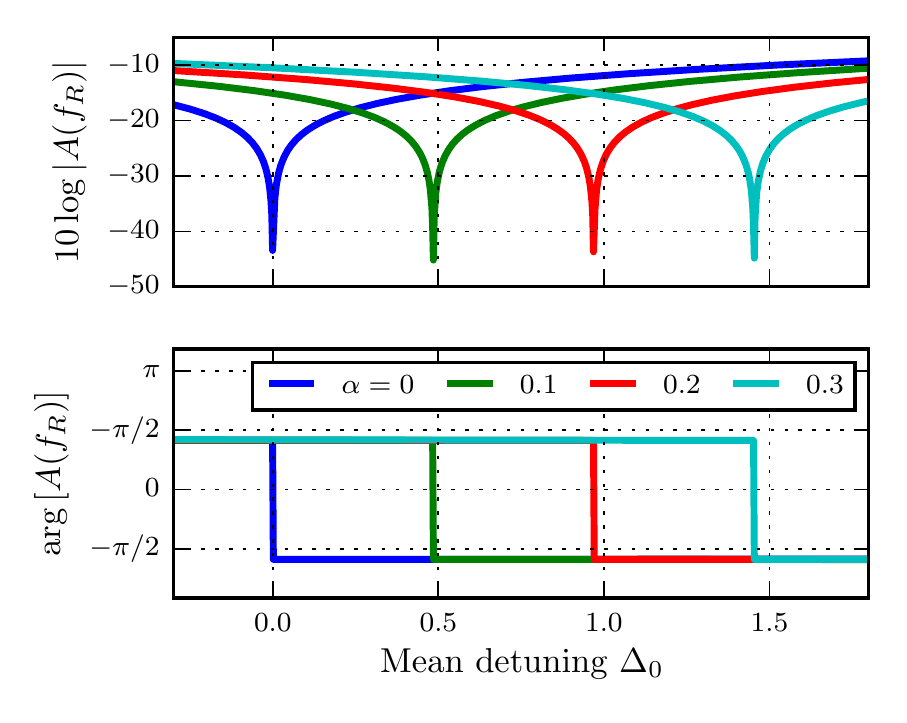}
	\caption{Maximum value of the transfer function $A(f_R)$ for amplitude response to a phase perturbation for different values of the mean detuning $\Delta_0$. One gets closer to zero response when approaching $\Delta_0=\Delta_m$, and a phase shift $\pi$ is observed when $\Delta_0$ crosses $\Delta_m$.}
	\label{fig:transfer}
\end{figure}
\par We introduce the phase modulation through the detuning, thus taking in account the FM modulation applied to the acousto-optic modulator. Therefore we consider $\Delta=\Delta_0+acos(2\pi f_{M}t)$.

The response of the system to this modulation according to the model is obtained by linearizing the equations around the equilibrium. We can then introduce the modulation as a small perturbation, and obtain the transfer function $A(f_M)$ for the output intensity. As expected, it features a resonant maximum at $f_R$, which suggests we should modulate at this particular frequency. The amplitude and phase of the maximum value $A(f_R)$ of this transfer function is plotted in Fig.~\ref{fig:transfer}. While for $\alpha=0$, the minimal (zero) response is obtained at $\Delta_0=0$, it is not anymore the case when $\alpha\neq 0$. The minimal response is shifted to a particular value $\Delta_m>0$, that depends strongly on $\alpha$. This value also corresponds to a phase jump for the transfer function. This confirms that the balance of the modulation sidebands in the spectrum is a good way to measure $\Delta_m$. 




The same reasoning that led to equation (\ref{eq:delta_m_inj}) also applies in this more complex case, resulting in the relation 

\begin{equation}
	\label{eq:min_cond}
	\Delta_m = \alpha \Gamma \left| \frac{\;\widehat{e_x}\;}{\;\widehat{e_y}\; }\right|
\end{equation}
similar to equation (\ref{eq:delta_m_inj}). 
For low injection level, one can consider that $\widehat{e}_{x,y}$ do not differ appreciably from their equilibrium values in the free-running regime, so that equation (\ref{eq:min_cond}) further simplifies to $\Delta_m = \alpha \Gamma$. In the general case $\widehat{e}_{x,y}$ have to be computed from the system equations (\ref{eq:model1}-\ref{eq:model3}), and inserted into equation (\ref{eq:min_cond}). This procedure leads to a 4th degree polynomial for $\Delta_m$, which can be solved for a single real value

\begin{equation}
	\label{eq:delta_m_comp}
	\Delta_m=\alpha\Gamma(1+f(\epsilon\Gamma,\beta,\eta))
\end{equation}

where $f$ is a correction function that cancels for $\Gamma=0$, and has the rather cumbersome expression to the first order in $\epsilon\Gamma$ :

\begin{equation}
	\label{eq:correctionfunc}
	\begin{split}
		& f(\epsilon\Gamma,\Omega,\eta)  = \biggl\{\Gamma \epsilon(\eta \Omega^{2} - 2\eta - 2 \Omega^{2}) - \Omega^{2}  \\ &
		+ \biggl(- 2 \Gamma^{2} \epsilon^{2} (- \eta^{2} \Omega^{4}/2 - \eta^{2} \Omega^{2} + \eta^{2} + 3 \eta \Omega^{4} + \eta \Omega^{2}   \\ &- 3 \Omega^{4} + \Omega^{2} ) + 2 \Gamma \epsilon \Omega^{2}  (- \eta \Omega^{2} + \eta + 2 \Omega^{2} ) + \Omega^{4} \biggl)^\frac{1}{2} \biggl \} \\ &\biggl/\biggl(3 \Gamma \epsilon ( -\eta \Omega^{2} +  \eta + 2 \Omega^{2}) + \Omega^{2}\biggl)
	\end{split}
\end{equation}


This shows that for higher injection level, the minimal response detuning is not simply $\alpha\Gamma$, but it depends on other parameters of the model.

\par We can now use equation (\ref{eq:delta_m_comp}) to extract the value of $\alpha$ from the measured results of Fig~\ref{fig:result}.
From each point of  Fig~\ref{fig:result}, a value of $\alpha$ can be estimated, so that the final result can be obtained simply by averaging :
\begin{equation}
	\alpha=\left<\frac{\Delta_m}{\Gamma(1+f(\epsilon\Gamma,\beta,\eta))}\right>
\end{equation}
To take in account that the the uncertainty on a single measurement $\delta\alpha$ is higher for low values of $\Gamma$, we use a weighted average with weights $1/\delta\alpha$, and we find a value of $\alpha = \alphavalue$. The uncertainty is computed from the precision of the frequency measurements, which we estimate to be around $8\mathrm{kHz}$, and also includes the uncertainties on $\beta$ and $\eta$. This leads to a satisfying reduced chi-squared value of 1.15.

\par In conclusion, a new method for characterizing $\alpha$ has been demonstrated through the measurement of the inherent small phase-amplitude coupling in a Nd\textsuperscript{3+}:YAG solid-state laser. It is based on frequency modulated injection, so that it completely frees the measure from pump modulation and associated strong thermally induced AM/FM coupling. The use of laser operating in a dual-frequency regime makes it easier to solve the problem of the mode-matching of master and slave lasers in injection setups, and circumvents the need of a stable reference laser for heterodyning.  

\par The physical origin of the phase-amplitude coupling in rare-earth active medium was left out of the scope of this letter, but we can already note that the value $\alpha = \alphavalue$ is quite close to the $0.25\pm 0.13$ reported for Nd\textsuperscript{3+}:YVO\textsubscript{4} in Ref.~\cite{fordell2005}, suggesting very little influence of the host crystal matrix on $\alpha$. Our method can be applied to Er\textsuperscript{3+}-doped bulk or fibered medium. For instance, this may give clues to the potential contribution of $\alpha$ to the AM/FM noise conversion process during low phase noise microwave or THz generation~\cite{quinlan2011,rolland2014a}.

\par The authors thank Marc Brunel for fruitful discussion, and Steve Bouhier and Ludovic Frein for technical support in electronics.




\bibliography{ao_alpha}

\end{document}